\documentclass[11pt]{article}

\usepackage[top=1in,bottom=1in,left=1in,right=1in]{geometry}
\usepackage{natbib}
\usepackage[unicode=true,bookmarks=true,bookmarksnumbered=false,bookmarksopen=false,breaklinks=false,pdfborder={0 0 1},backref=false,colorlinks=true]{hyperref}
\hypersetup{citecolor=blue}
\usepackage{url}

\usepackage{amsmath}
\usepackage{mathrsfs}
\usepackage{amssymb}
\usepackage{amsthm}
\usepackage[normalem]{ulem}

\newtheorem{proposition}{Proposition}

\newtheorem{definition}{Definition}

\usepackage{float}
\usepackage{rotfloat}

\floatstyle{ruled}
\newfloat{algorithm}{tpb}{loa}
\providecommand{\algorithmname}{Algorithm}
\floatname{algorithm}{\protect\algorithmname}

\usepackage{graphicx}
\usepackage[caption=false]{subfig} 

\usepackage{array}
\newcolumntype{L}[1]{>{\raggedright\let\newline\\\arraybackslash\hspace{0pt}}m{#1}}
\newcolumntype{C}[1]{>{\centering\let\newline\\\arraybackslash\hspace{0pt}}m{#1}}
\newcolumntype{R}[1]{>{\raggedleft\let\newline\\\arraybackslash\hspace{0pt}}m{#1}}

\usepackage{rotating}
\usepackage{multirow}

\usepackage{tikz}
\usetikzlibrary{shapes,arrows,backgrounds,calc,positioning,fit,petri,plotmarks}
\usetikzlibrary{arrows}

\usepackage{enumerate}
\usepackage{paralist}

\usepackage{chngcntr}         

\newcommand*{\affaddr}[1]{#1} 
\newcommand*{\affmark}[1][*]{\textsuperscript{#1}}

\title{$B$-Value and Empirical Equivalence Bound: A New Procedure of Hypothesis Testing}

\author{%
    Yi Zhao\affmark[1], Brian S. Caffo\affmark[2] and Joshua B. Ewen\affmark[3] \\
    \affaddr{\affmark[1]Department of Biostatistics, Indiana University School of Medicine} \\
    \affaddr{\affmark[2]Department of Biostatistics, Johns Hopkins Bloomberg School of Public Health} \\
    \affaddr{\affmark[3]Kennedy Krieger Institute and Johns Hopkins University School of Medicine} \\
}

\date{}

\begin{document}

\maketitle

\thispagestyle{empty}

\begin{abstract}
 In this study, we propose a two-stage procedure for hypothesis testing, where the first stage is conventional hypothesis testing and the second is an equivalence testing procedure using an introduced Empirical Equivalence Bound. In 2016, the American Statistical Association released a policy statement on $P$-values to clarify the proper use and interpretation in response to the criticism of \emph{reproducibility} and \emph{replicability} in scientific findings. A recent solution to improve reproducibility and transparency in statistical hypothesis testing is to integrate $P$-values (or confidence intervals) with practical or scientific significance. Similar ideas have  been proposed via the equivalence test, where the goal is to infer equality under a presumption (null) of inequality of parameters. However, in these testing procedures, the definition of scientific significance/equivalence can be subjective. To circumvent this drawback, we introduce a $B$-value and the Empirical Equivalence Bound, which are both estimated from the data. Performing a second-stage equivalence test, our procedure offers an opportunity to correct for false positive discoveries and improve the reproducibility in findings across studies.
\end{abstract}



\clearpage
\setcounter{page}{1}

\section{Introduction}
\label{sec:intro}

Confidence intervals (CIs) are widely used for statistical inference. In a traditional two-sample comparative study one uses a $100(1-\alpha)\%$ confidence interval of the difference to draw an inference. Via a hypothesis test of equal means (say), when the $100(1-\alpha)\%$ CI does not cover zero, it is concluded that there exists evidence of a difference between the two groups (controlling the Type I error rate at $\alpha$). When the CI does not cover zero, it is concluded that there is insufficient evidence to suggest a difference. However, as is often emphasized in introductory courses, this insufficient evidence of difference does not imply equivalence, absence of evidence is not evidence of absence, so to speak. To test for equivalence, an equivalence test is performed where the typical roles of the null and alternative are reversed and explicit bounds implying equivalence are specified. If the equivalence test is implemented in a \emph{post hoc} sense, i.e. if the confidence interval was verified to include zero prior to investigating equivalence, special care must be taken to avoid erroneous conclusions~\citep{hoenig2001abuse}. 

\begin{figure}
    \begin{center}
        \subfloat[\label{subfig:conclusion_NRej}$0\in\text{confidence interval}$]{\includegraphics[width=0.45\textwidth,page=9,trim=1cm 0cm 1cm 0cm,clip]{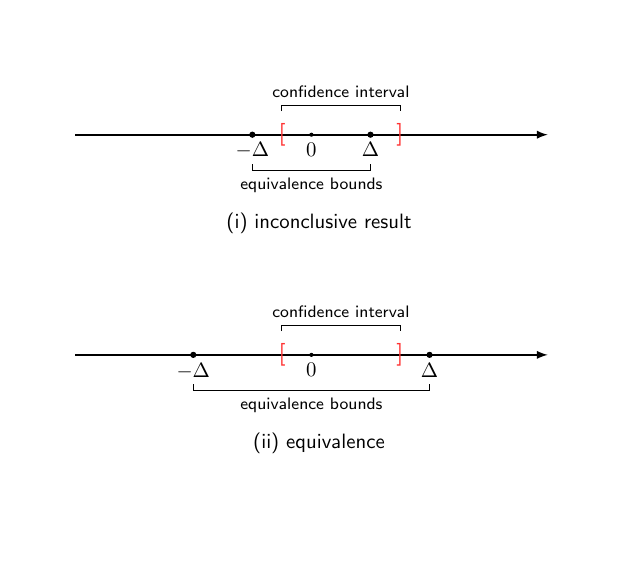}}
        \enskip{}
        \subfloat[\label{subfig:conclusion_Rej}$0\notin\text{confidence interval}$]{\includegraphics[width=0.45\textwidth,page=8,trim=1cm 0cm 1cm 0cm,clip]{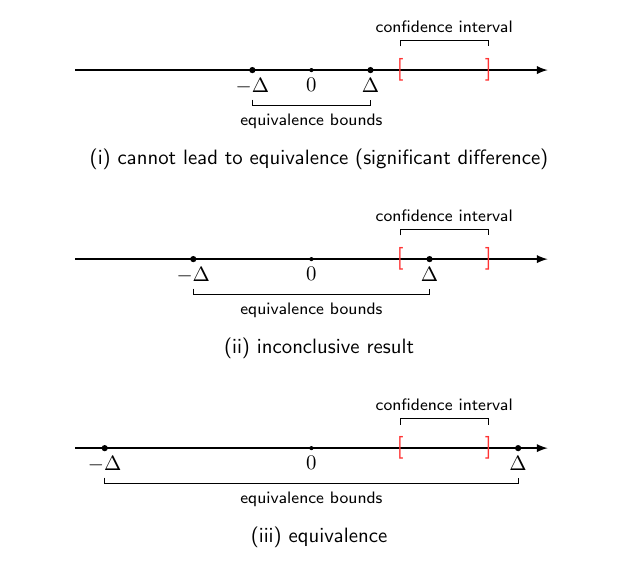}}
    \end{center}
    \caption{\label{fig:conclusion_type}Conclusion of an equivalence test with different prespecified equivalence bounds when (a) the confidence interval covers zero and (b) the confidence interval does not cover zero. Two possible conclusions under scenario (a): (i) inconclusive result and (ii) equivalence. Three possible conclusions under scenario (b): (i) significant difference, (ii) inconclusive result, and (iii) equivalence.}
\end{figure}

One critical concern about equivalence tests is the choice of equivalence bounds. In an equivalence study one rejects the null hypothesis of inequivalence if a $100(1-2\alpha)\%$ interval is entirely contained within the pre-specified equivalence bounds. The bounds, typically symmetric around zero, are chosen to represent a trivial difference so that a true difference less than the bounds is considered equivalent. For example, Figure~\ref{fig:conclusion_type} presents  possible conclusions with different choices of equivalence bounds. Of course, with narrow bounds, the equivalence test is less likely to conclude equivalence. 

In this study, we introduce the concepts of the $B$-value and the Empirical Equivalence Bound: the minimum equivalence bound in principle that leads to equivalence when equivalence is true. The $B$-value is analogous to the attained significance level interpretation of a $P$-value. That is, the attained significance level is the smallest type I error rate for which one would reject the null hypothesis. The $B$-value is the smallest symmetric equivalence bound for which one would reject in a test of equivalence. This is inherently useful when one wants to test equivalence, but does not have a natural bound to work with. By reporting the $B$-value, the reader can easily apply whatever bound they choose, not unlike how a reader can easily employ any error rate they choose if $P$-values are reported.

Equivalence test procedures have long been studied to examine the equivalence of two drug formulations. \citet{westlake1972use,westlake1976symmetrical} proposed the use of symmetric confidence intervals in lieu of conventional confidence intervals in bioequivalence trials. These symmetric confidence intervals decrease the effective length of the interval while increasing the confidence coefficient. Here, the effective length of a confidence interval $[L,U]$ is not $U-L$, but rather $2\max\{|L|,|U|\}$. \citet{anderson1983new} and \citet{hauck1984new} introduced $t$-test procedures that were shown to be more powerful than the symmetric/shortest confidence interval approach when testing for equivalence. \citet{schuirmann1987comparison} considered a two one-sided tests procedure. Compared with the \emph{power} method by \citet{hauck1984new}, the two one-sided tests procedure demonstrates superior properties. Hybridizing the power method and the two one-sided tests procedure, \citet{munk1993improvement} corrected the inflated Type I error rate in the power method. \citet{liu1990confidence}, \citet{hsu1994confidence} and \citet{seaman1998equivalence} proposed the use of $100(1-2\alpha)\%$ CIs to construct the so-called \emph{equivalence confidence interval}, which is also symmetric about zero, and showed that the equivalence confidence interval can lead to a more powerful test, as the effective length of the interval is smaller. \citet{seaman1998equivalence} also suggested a sequential method to assess the difference in two means. The method starts with a conventional two-sample $t$-test. If the null hypothesis cannot be rejected, one can proceed to the equivalence test comparing the equivalence confidence interval with the nominal equivalence interval. For other extensions to the equivalence test procedure, one can refer to the discussion in \citet{seaman1998equivalence}. When the significance of the $t$-test is not obtained, various studies advocate for a post-experiment power calculation. However, this approach sits on an inappropriate statistical hypothesis~\citep{hauck1984new,schuirmann1987comparison} and suffers from fatal logical flaws, as discussed in \citet{hoenig2001abuse}. As an alternative, a confidence interval or equivalence test was suggested in \citet{hoenig2001abuse}.

In this study, we follow the sequential method recommended by \citet{seaman1998equivalence}, while further studying the properties of the equivalence confidence interval and addressing the issue of how to determine a nominal equivalence bound. In Section~\ref{sec:method}, we introduce the concept of $B$-value, which is defined as the maximum magnitude of the $100(1-2\alpha)\%$ CI bounds. The $B$-value is the smallest symmetric equivalence bound for which one would conclude equivalence. We derive the distribution of the $B$-value, as well as the conditional distribution based on the hypothesis testing result in the conventional two-sample $t$-test. Based on these distributions, we then introduce the Empirical Equivalence Bound (EEB), which can be used for equivalence tests. A two-stage testing procedure to compare two group means is suggested. This data-driven procedure requires no prior knowledge as to what level the two groups are equivalent. In Section~\ref{sec:plant}, we apply our proposed procedure to the Plant Growth Data~\citep{dobson1983introduction} available in the open source software \textsf{R}~\citep{Rsoftware}. Section~\ref{sec:discussion} gives a summary and discussion.

\section{$B$-Value and the Empirical Equivalence Bound}
\label{sec:method}

\subsection{Formulation}
\label{sub:method_formulation}

Consider a two-sample $t$-test setting with hypotheses
\begin{equation}
\label{eq:normalHypotheses}
    \mathrm{H}_{0}:\delta=0 \quad ~~\mbox{versus}~~ \quad \mathrm{H}_{1}:\delta\neq 0,
\end{equation}
where $\delta=\mu_{1}-\mu_{2}$ is the difference of two population averages. A standard procedure for performing hypothesis testing is to construct the confidence interval. Let $[L_{0},U_{0}]$ denote the $100(1-\alpha)\%$ confidence interval, where
\[
    L_{0}=\hat{\delta}-t_{\nu,1-\alpha/2}S, \quad U_{0}=\hat{\delta}+t_{\nu,1-\alpha/2}S,
\]
$\hat{\delta}=\bar{x}_{1}-\bar{x}_{2}$ is an estimate of $\delta$ with $\bar{x}_{1}$ and $\bar{x}_{2}$ as the sample average of the two groups; $t_{\nu,1-\alpha/2}$ is the $100(1-\alpha/2)\%$ quantile of a $t$-distribution with degrees of freedom $\nu$; $S$ is the pooled standard error under the assumption of constant variances across groups:
\begin{equation}
    S=\sqrt{\frac{1}{n_{1}}+\frac{1}{n_{2}}}\times\sqrt{\frac{(n_{1}-1)S_{1}^{2}+(n_{2}-1)S_{2}^{2}}{n_{1}+n_{2}-2}},
\end{equation}
where $S_{1}^{2}$ and $S_{2}^{2}$ are the sample variance of the two groups, and $n_{1}$ and $n_{2}$ are the sample sizes. Hypothesis test \eqref{eq:normalHypotheses} is based on whether $[L_{0},U_{0}]$ covers zero. However, when $0\in[L_{0},U_{0}]$, via normal testing logic, one cannot directly conclude equivalence of the two groups. 

As suggested in \citet{seaman1998equivalence}, in order to evaluate equivalence, an equivalence test can be conducted. Later, we discuss the implications of testing \eqref{eq:normalHypotheses} test then performing equivalence testing if the result is a failure to reject and, for completeness, also consider testing for equivalence when it is rejected. For now, assume the equivalence test is the only test performed.

In equivalence testing, one is testing the hypotheses
\begin{equation}
\label{eq:equivalenceHypotheses}
    \mathrm{H}_{3}: |\delta| \geq \Delta \quad ~~\mbox{versus}~~ \quad \mathrm{H}_{4}: |\delta| < \Delta
\end{equation}
where $\Delta$ is a pre-specified equivalence bound. The alternative hypothesis, $|\delta| < \Delta$ represents equivalence in the sense that $\Delta$ is chosen to represent a trivially small difference given the context. Here, 
instead of using the $100(1-\alpha)\%$ confidence interval, a $100(1-2\alpha)\%$ confidence interval is formulated~\citep{seaman1998equivalence}, denoted as $[L,U]$, where
\begin{equation}
    L=\hat{\delta}-t_{\nu,1-\alpha}S, \quad U=\hat{\delta}+t_{\nu,1-\alpha}S,
\end{equation}
and $t_{\nu,1-\alpha}$ is the $100(1-\alpha)\%$ quantile of a $t$-distribution with degrees of freedom $\nu$. The classic equivalence test is to compare this interval with  predetermined equivalence bounds. If the interval lies entirely within the bounds, the null hypothesis is rejected (equivalence concluded, see Figure~\ref{fig:conclusion_type}). Otherwise, the null hypothesis is not rejected and there is insufficient evidence to conclude equivalence. It is important to emphasize that the conclusion is subject to the choice of $\Delta$. In this study, we propose a procedure in which the equivalence bound is derived from the data, and we call it the \emph{Empirical Equivalence Bound} (EEB).

To obtain the EEB, we first introduce the \emph{$B$-value}, $B=\max\{|L|,|U|\}$. If one takes nothing else from this manuscript, consider that the $B$-value is useful to report in the sense that if $B < \Delta$ one rejects and concludes equivalence. This is analogous to the attained significance level where one rejects if $P<\alpha$. Since it is common to want to test equivalence, but not be in possession of meangingful bounds, reporting $B$ is useful as a reader can then perform the test on whatever bounds they believe are most relevant.

In this manuscript we investigate the properties of $B$. Specifically, we derive the distribution of $B$, as well as the conditional distribution of $B$ given the test result of the first-step hypothesis testing \eqref{eq:normalHypotheses}; that is, conditional on $0\in[L_{0},U_{0}]$ or $0\notin[L_{0},U_{0}]$. This is useful as it is common to investigate equivalence after failing to reject classic hypotheses. However, it is well understood that ignoring the first rejection leads to erroneous conclusions~\citep{seaman1998equivalence,hoenig2001abuse}.

We also consider the distribution of $B$ given $0\notin[L_{0},U_{0}]$. That is, considering testing equivalence, hypotheses \eqref{eq:equivalenceHypotheses}, after having rejected the null hypothesis from \eqref{eq:normalHypotheses}. This is done primarily for completeness. However, it's possible that a researcher might want to investigate potential triviality of a rejection of the traditional hypothesis test.

The most related study is \citet{seaman1998equivalence}, where $[-B,B]$ was referred to as the \emph{equivalence confidence interval} which was recommended for use when the effect is small. In Definition~\ref{def:dist_B}, we focus on the scenario when the true $\delta$ is zero, exact equivalence. In Section~\ref{appendix:sec:dist_B} of the supplementary material, we provide the (conditional) distribution of $B$ under the general scenario of $\delta\in\mathbb{R}$. 
\begin{definition}\label{def:dist_B}
    Consider the parameter of interest $\delta=\mu_{1}-\mu_{2}$ and the hypothesis testing problem $\mathrm{H}_{0}:\delta=0$. The test statistic $t=(\hat{\delta}-\delta)/S$ follows a Student's $t$-distribution with degrees of freedom $\nu$, where $\hat{\delta}$ is the estimate of $\delta$ and $S$ is the standard error. Denote $[L_{0},U_{0}]$ and $[L,U]$ as the $100(1-\alpha)\%$ and $100(1-2\alpha)\%$ confidence intervals associated with the estimate $\hat{\delta}$, respectively, where $L_{0}=\hat{\delta}-t_{\nu,1-\alpha/2}S$, $U_{0}=\hat{\delta}+t_{\nu,1-\alpha/2}S$, $L=\hat{\delta}-t_{\nu,1-\alpha}S$, $U=\hat{\delta}+t_{\nu,1-\alpha}S$ and $t_{\nu,q}$ is the $100q\%$ quantile of a Student's $t$-distribution with degrees of freedom $\nu$. Define the \textbf{$B$-value} as $B=\max\{|L|,|U|\}$. When the true $\delta=0$, 
    \begin{enumerate}[(1)]
        \item the cumulative distribution function of $B$ is:
            \begin{equation*}
                F_{B}(b~|~\mathrm{H}_{0})=\begin{cases}
                        0 & \text{if } b<St_{\nu,1-\alpha} \\
                        2F_{t}(b/S-t_{\nu,1-\alpha};\nu)-1 & \text{if } b\geq St_{\nu,1-\alpha}
                \end{cases};
            \end{equation*}
        \item the conditional cumulative distribution function of $B$ given $0\in[L_{0},U_{0}]$ is:
            \begin{equation*}
                F_{B}(b~|~0\in[L_{0},U_{0}],\mathrm{H}_{0})=\begin{cases}
                    0 & \text{if } b<St_{\nu,1-\alpha} \\
                    \left\{2F_{t}(b/S-t_{\nu,1-\alpha};\nu)-1\right\}/(1-\alpha) & \text{if } St_{\nu,1-\alpha}\leq b<S(t_{\nu,1-\alpha}+t_{\nu,1-\alpha/2}) \\
                    1 & \text{if } b\geq S(t_{\nu,1-\alpha}+t_{\nu,1-\alpha/2})
                \end{cases};
            \end{equation*}
        \item the conditional cumulative distribution function of $B$ given $0\notin[L_{0},U_{0}]$ is:
            \begin{equation*}
                F_{B}(b~|~0\notin[L_{0},U_{0}],\mathrm{H}_{0})=\begin{cases}
                    0 & \text{if } b<S(t_{\nu,1-\alpha}+t_{\nu,1-\alpha/2}) \\
                    \left\{F_{t}(b/S-t_{\nu,1-\alpha};\nu)-(1-\alpha/2)\right\}/(\alpha/2) & \text{if } b\geq S(t_{\nu,1-\alpha}+t_{\nu,1-\alpha/2})
                \end{cases},
            \end{equation*}
    \end{enumerate}
    where $F_{t}(\cdot;\nu)$ is the cumulative distribution function of a Student's $t$-distribution with degrees of freedom $\nu$.
\end{definition}

The \emph{Empirical Equivalence Bound} is defined as the bound of an equivalence test such that when the true population difference is exactly zero, an equivalence test rejects with probability $\beta$.
\begin{definition}[Empirical Equivalence Bound]\label{def:EEB}
    Assume that the parameter of interest is the difference in the  population means, $\delta=\mu_{1}-\mu_{2}$. Consider a hypothesis testing problem $\mathrm{H}_{0}:\delta=0$ of level $\alpha$. For a given $\beta\in(0,1)$, under test result $C$, the Empirical Equivalence Bound at level $\beta$ is defined as:
    \begin{equation}
        \mathrm{EEB}_{\alpha}(\beta~|~C) = \inf_{b\in[0,\infty]} \left\{b:F_{B}(b~|~C,\mathrm{H}_{0})\geq \beta\right\}.
    \end{equation}
    Here, $C\in\{\emptyset,0\in[L_{0},U_{0}],0\notin[L_{0},U_{0}]\}$ denotes the status of the hypothesis test, and $F_{B}(\cdot~|~C,\mathrm{H}_{0})$ is the conditional cumulative distribution function of $B$. If the test result is unknown, $C=\emptyset$, and $F_{B}(\cdot~|~\emptyset,\mathrm{H}_{0})=F_{B}(\cdot~|~\mathrm{H}_{0})$ is the marginal distribution of $B$.
\end{definition}

Under the condition $0\in[L_{0},U_{0}]$ or $0\notin[L_{0},U_{0}]$, the $\mathrm{EEB}(\beta)$ defined in Definition~\ref{def:EEB} has the following explicit form:
\begin{eqnarray*}
    \mathrm{EEB}_{\alpha}(\beta~|~0\in[L_{0},U_{0}]) &=& S\left\{F_{t}^{-1}\left(\frac{\beta(1-\alpha)+1}{2};\nu\right)+t_{\nu,1-\alpha}\right\}, \label{eq:EEB_NRej} \\
    \mathrm{EEB}_{\alpha}(\beta~|~0\notin[L_{0},U_{0}]) &=& S\left\{F_{t}^{-1}\left(1-\frac{\alpha(1-\beta)}{2};\nu\right)+t_{\nu,1-\alpha}\right\}, \label{eq:EEB_Reg}
\end{eqnarray*}
where $F_{t}^{-1}(\cdot;\nu)$ is the inverse cumulative distribution function of a Student's $t$-distribution with degrees of freedom $\nu$.

\subsection{Properties of the EEB}
\label{sub:property}

The EEB has the following properties:
\begin{enumerate}[(i)]
    \item For fixed $\alpha$, $\mathrm{EEB}_{\alpha}(\beta~|~C)$ is a non-decreasing function of $\beta$;
    \item For fixed $\beta$, $\mathrm{EEB}_{\alpha}(\beta~|~C)$ is a non-increasing function of $\alpha$.
\end{enumerate}
The proof of these two properties is straightforward following the fact that the cumulative density function is right continuous and non-decreasing. With the same significance level in the first-step $t$-test, it requires the equivalence interval to be wider for a higher confidence in the second-step equivalence test. On the other hand, when the level in the second-step equivalence test is fixed, a higher confidence level in the first step (lower $\alpha$) demands a wider equivalence interval as well.

The following proposition presents the relationship between the three cumulative density functions in Definition~\ref{def:dist_B} and the correspondingly defined EEB.
\begin{proposition}\label{prop:EEB}
    Consider the three distributions of $B$ defined in Definition~\ref{def:dist_B}. For any $b$, we have
    \[
        F_{B}(b~|~0\notin[L_{0},U_{0}],\mathrm{H}_{0})\leq F_{B}(b~|~\mathrm{H}_{0})\leq F_{B}(b~|~0\in[L_{0},U_{0}],\mathrm{H}_{0}).
    \]
    Therefore, for fixed $\alpha$ and $\beta$, the EEBs have the following relationship
    \[
        \mathrm{EEB}_{\alpha}(\beta~|~0\notin[L_{0},U_{0}])\geq\mathrm{EEB}_{\alpha}(\beta)\geq\mathrm{EEB}_{\alpha}(\beta~|~0\in[L_{0},U_{0}]),
    \]
    where $\mathrm{EEB}_{\alpha}(\beta)=\mathrm{EEB}_{\alpha}(\beta~|~\emptyset)$.
\end{proposition}
Proposition~\ref{prop:EEB} demonstrates that, given the result from the two-sample $t$-test, the conditional EEB appropriately shrinks/expands the equivalence interval in the equivalence test. For example, if the two-sample $t$-test does not reject the null, the $B$-value, as well as the conditional EEB, shrinks toward zero. If the test rejects the null even though the true parameter is zero, then a wider equivalence interval is required to correct for this false positive, which is achieved with a greater EEB value. In other words, the EEB allows one to interpret equivalence post testing.

\subsection{A two-stage testing procedure}
\label{sub:testproc}

\begin{figure}
    \begin{center}
        \includegraphics[width=0.9\textwidth,page=10,trim=0cm 1cm 0cm 1cm,clip]{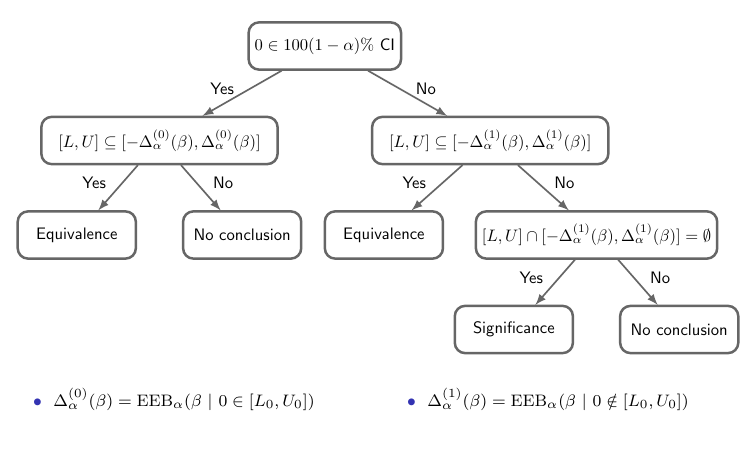}
    \end{center}
    \caption{\label{fig:testproc}A two-stage testing procedure comparing two means.}
\end{figure}

Using the defined EEB, we propose a two-stage testing procedure when comparing two means. The procedure is summarized in Figure~\ref{fig:testproc}. The first stage is the conventional two-sample $t$-test. Based on whether the $100(1-\alpha)\%$ confidence interval covers zero, we calculate the conditional EEB at level $\beta$ denoted as $\Delta_{\alpha}^{(r)}(\beta)$, where $r=0$ if $0\in 100(1-\alpha)\%$ CI and $r=1$ otherwise. The second stage compares the $100(1-2\alpha)\%$ CI (denoted by $[L,U]$) and the $\beta$-level empirical equivalence interval (denoted by $[-\Delta_{\alpha}^{(r)}(\beta),\Delta_{\alpha}^{(r)}(\beta)]$ for $r=0,1$). When the first stage result is $0\in 100(1-\alpha)\%$ CI, if the $100(1-2\alpha)\%$ CI is fully contained in the empirical equivalence interval, there is sufficient evidence to conclude equivalence of the two groups. Otherwise, no confirmatory conclusion can be achieved. When the first stage result is $0\notin 100(1-\alpha)\%$ CI, if the empirical equivalence interval covers the $100(1-2\alpha)\%$ CI, one can conclude that the two means are actually equivalent and the second-stage equivalence test corrects the false positive discovery in the first stage. If there is overlap between the two intervals, there is no confirmatory conclusion. For sufficient high $\beta$ level, if there is no overlap between the two intervals, this can be seen as a confirmation of the significant finding in the first stage.

\subsection{An example}
\label{sub:example}

In this section, we use simulated examples to elaborate the definition and the properties of the $B$-value and the EEB. Assuming that the true $\delta$ parameter is zero, and we generate two-sample data from equivalent normal distributions.  The estimate of $\delta$, standard error ($S$) and the $100(1-\alpha)\%$ confidence interval are then attained. By examining if the $100(1-\alpha)\%$ confidence interval covers zero, with a designated level $\beta$, we then calculate the EEB under the corresponding condition and compare with the $100(1-2\alpha)\%$ confidence interval.

\begin{figure}
    \begin{center}
        \subfloat[\label{subfig:example_eg1_marginal}$C=\emptyset$]{\includegraphics[width=0.3\textwidth,page=1]{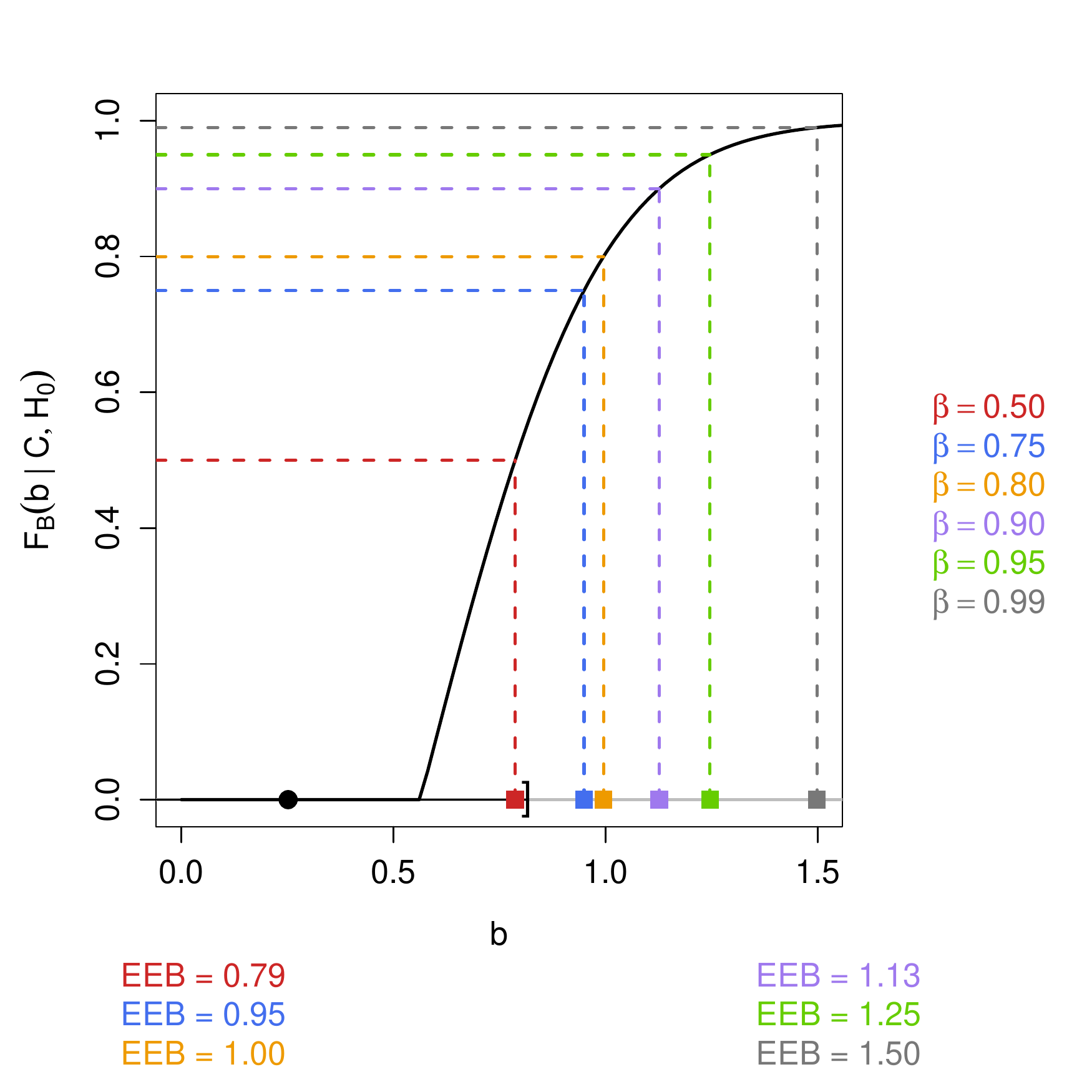}}
        \enskip{}
        \subfloat[\label{subfig:example_eg1_conditional}$C=0\in{[L_{0},U_{0}]}$]{\includegraphics[width=0.3\textwidth,page=2]{190603_eg_EEB_conNRej}}
        \enskip{}
        \subfloat[\label{subfig:example_eg2_conditional}$C=0\notin{[L_{0},U_{0}]}$]{\includegraphics[width=0.3\textwidth,page=2]{190603_eg_EEB_conRej}}
    \end{center}
    \caption{\label{fig:example}Marginal and conditional distribution of the $B$-value and the corresponding EEB at various $\beta$ levels in the example.}
\end{figure}

\begin{figure}
    \begin{center}
        \subfloat[\label{subfig:example_distcomp}distribution of $B$-value]{\includegraphics[width=0.45\textwidth,page=1]{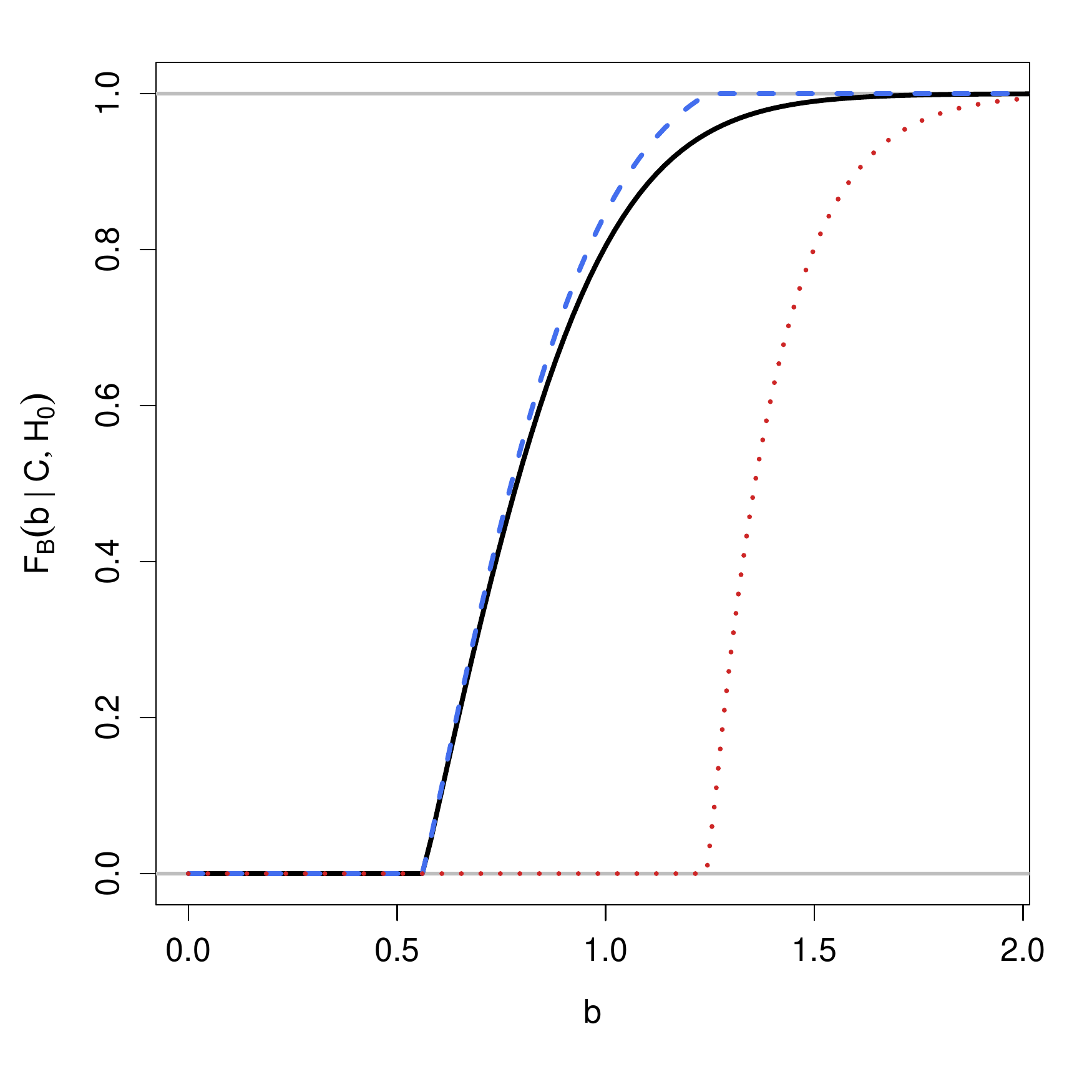}}
        \enskip{}
        \subfloat[\label{subfig:example_EEBcomp}$\mathrm{EEB}_{\alpha}(\beta~|~C)$]{\includegraphics[width=0.45\textwidth,page=1]{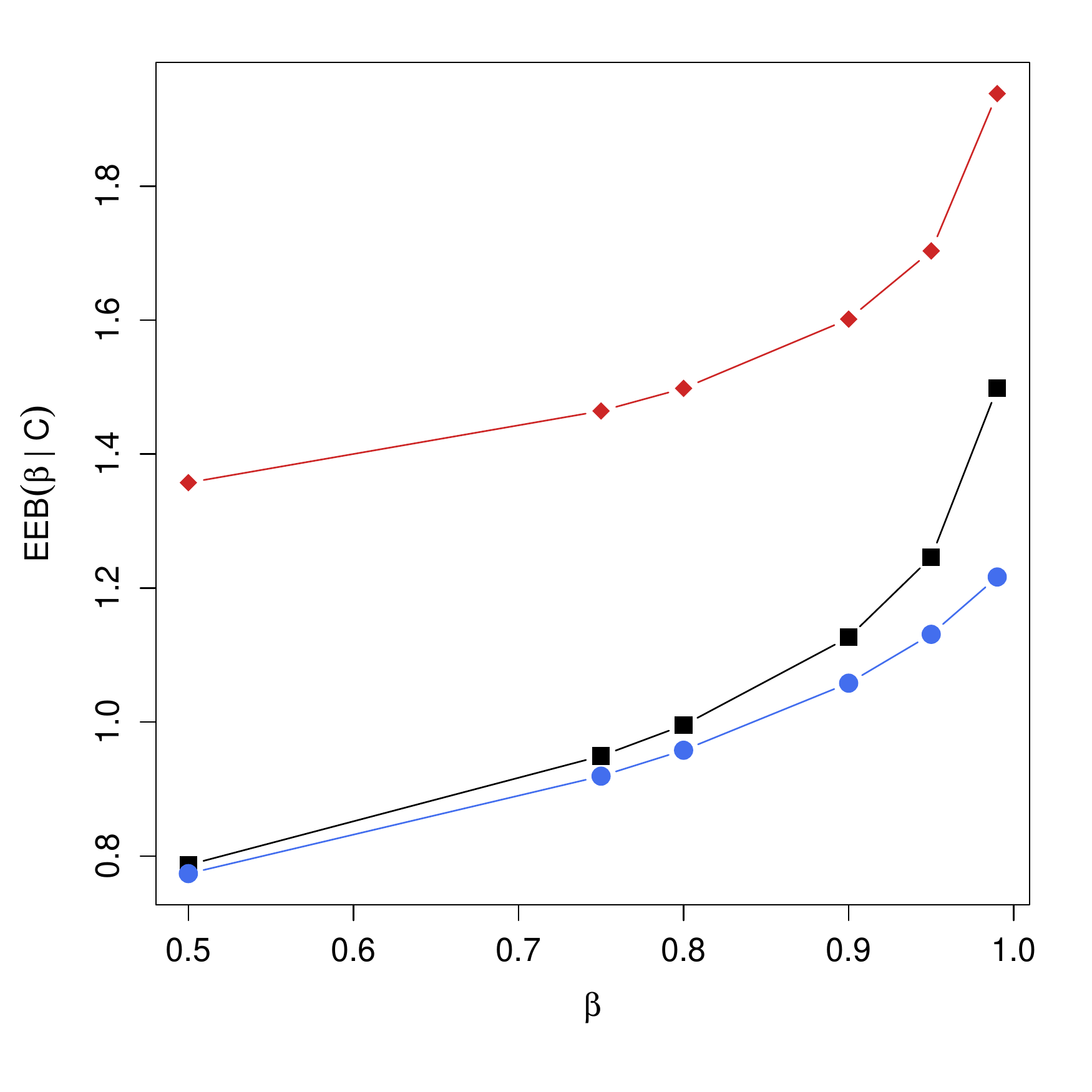}}

        \includegraphics[width=0.7\textwidth]{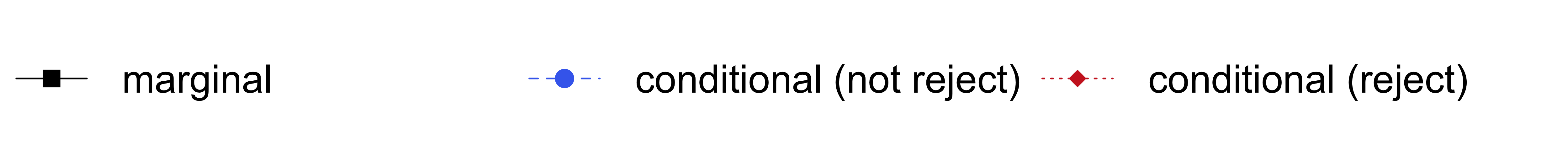}
    \end{center}
    \caption{\label{fig:example_EEBcomp}(a) The marginal and conditional distribution of the $B$-value and (b) the marginal and conditional EEB at various $\beta$ levels in the example.}
\end{figure}

For example, with the sample sizes $n_{1}=n_{2}=10$ (and thus $\nu=18$), we generate data for both groups from a standard normal distribution (i.e. $\delta=0$). We simulate data for both conditions: (i) $0\in[L_{0},U_{0}]$ and (ii) $0\notin[L_{0},U_{0}]$. Here, we compare between the two conditions, we keep the standard error the same, and generate $\hat{\delta}$ from the sample mean distribution. The statistics are presented in Table~\ref{table:example}. Figure~\ref{subfig:example_eg1_marginal} presents the marginal distribution of the $B$-value, and Figures~\ref{subfig:example_eg1_conditional} and \ref{subfig:example_eg2_conditional} present the conditional distribution, together with the EEB at various $\beta$-values. Figures~\ref{subfig:example_distcomp} and \ref{subfig:example_EEBcomp} compare the three distributions and the EEB values and to verify Proposition~\ref{prop:EEB}. Under condition (i), for given $\beta$, the knowledge of not rejecting the null in the two-sample $t$-test decreases the EEB value, making it more stringent in the equivalence test. Under condition (ii), with a bigger conditional EEB, the equivalence interval is wider, so that performing an equivalence test may help rectify the false positive finding in the two-sample $t$-test. Therefore, conditional on the result from the two-sample $t$-test, the conditional EEB improves the performance of the second-step equivalence test. Figure~\ref{fig:example_EEBcomp} also demonstrates one property of the EEB that for fixed significance level $\alpha$, the EEB is a non-decreasing function of $\beta$.

\begin{table}
    \caption{\label{table:example}Statistics in the two-sample $t$-test using the simulated data under conditions (i) and (ii) in the example.}
    \begin{center}
        \begin{tabular}{c c c c c}
            \hline
            Condition & $\hat{\delta}$ & $S$ & $[L_{0},U_{0}]$ & $[L,U]$ \\
            \hline
            (i) & $0.262$ & $0.325$ & $[-0.431,0.934]$ & $[-0.311,0.815]$  \\
            (ii) & $0.685$ & $0.325$ & $[0.002,1.367]$ & $[0.121,1.248]$ \\
            \hline
        \end{tabular}
    \end{center}
\end{table}


\subsection{Generalization to $z$-test}

Many parameter estimates enjoy asymptotic Gaussianity, such as method of moments and maximum likelihood estimators. For these estimators, hypothesis testing can be conducted through a $z$-test, where the $z$-score is calculated based on the asymptotic distribution. For a $z$-test, one can replace the $t$-distribution quantiles and cumulative distribution function in Section~\ref{sub:method_formulation} with the corresponding quantities from the standard normal distribution and all the results follow.





\section{Plant Growth Data}
\label{sec:plant}

\begin{figure}
    \begin{center}
        \subfloat[Boxplot of data]{\includegraphics[width=0.3\textwidth]{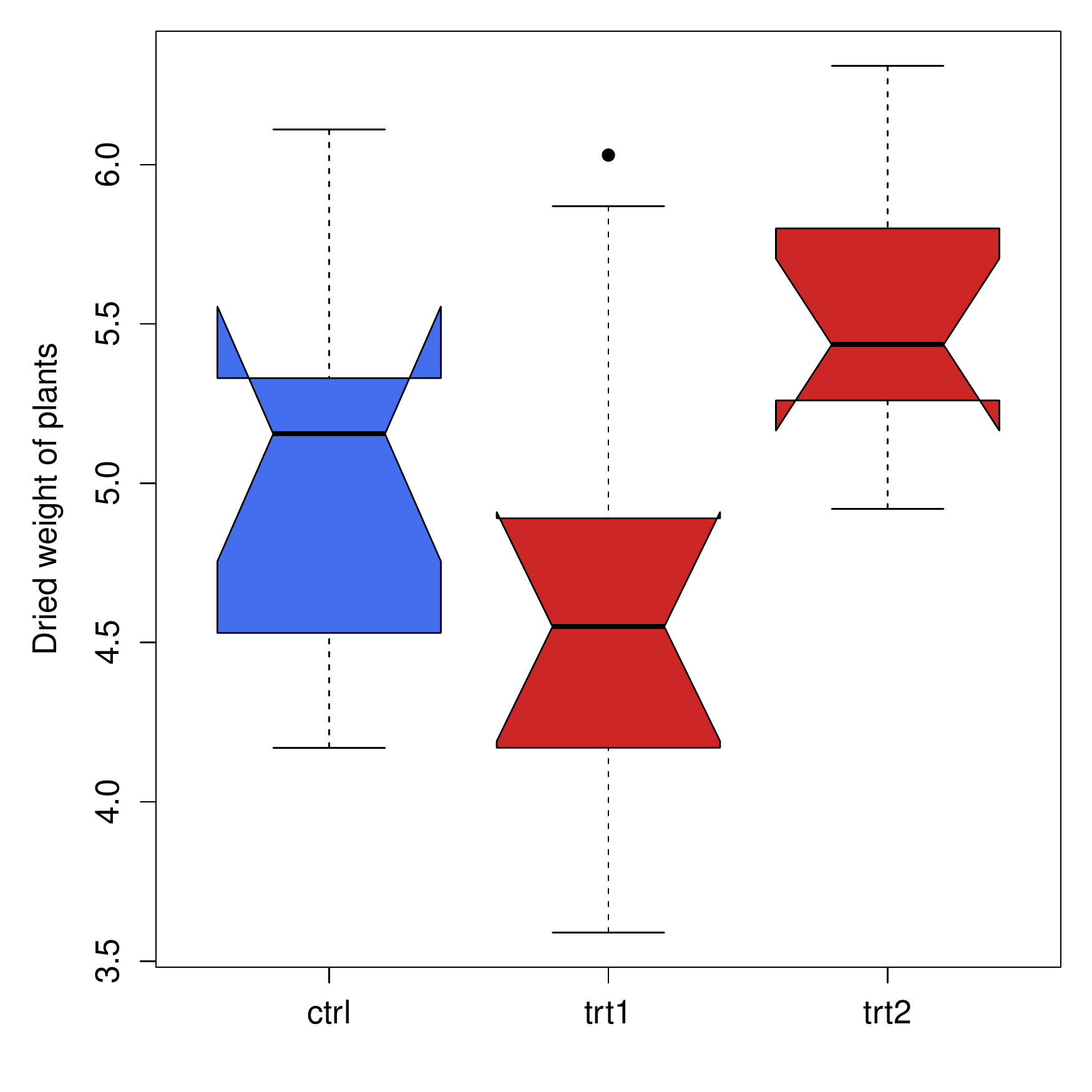}}
        \enskip{}
        \subfloat[Histogram and density of control (blue) and treatment 1 (red).]{\includegraphics[width=0.3\textwidth,page=1]{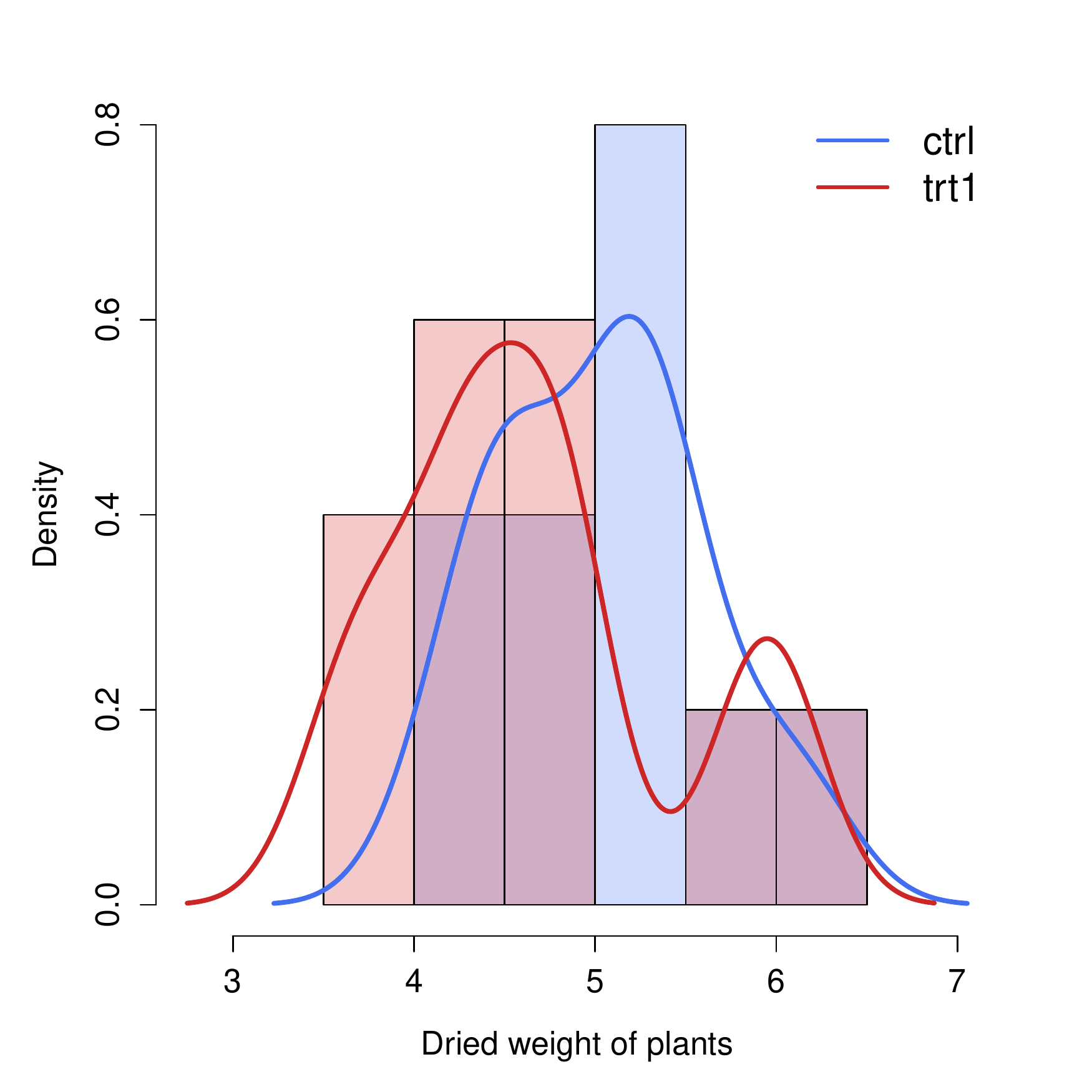}}
        \enskip{}
        \subfloat[Histogram and density of control (blue) and treatment 2 (red).]{\includegraphics[width=0.3\textwidth,page=2]{190603_PG_hist}}
    \end{center}
    \caption{\label{fig:plant_dist}(a) Boxplot and (b)\&(c) histogram and density function of the Plant Growth data for each group. ctrl: control group (in blue); trt1: treatment 1 (in red); trt2: treatment 2 (in red).}
\end{figure}

\begin{figure}
    \begin{center}
        \subfloat[marginal EEB of $\text{trt1}-\text{ctrl}$]{\includegraphics[width=0.45\textwidth]{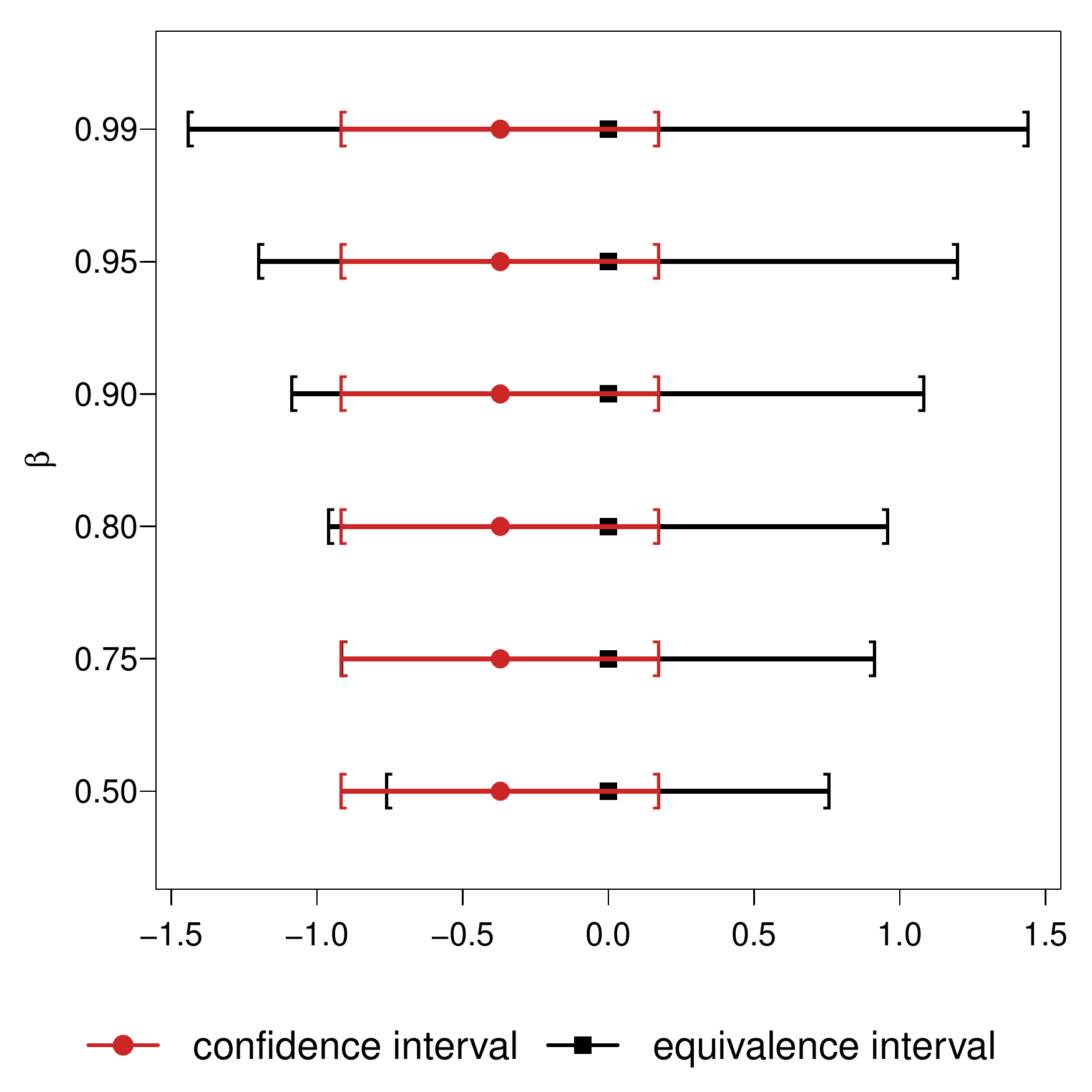}}
        \enskip{}
        \subfloat[conditional EEB of $\text{trt1}-\text{ctrl}$]{\includegraphics[width=0.45\textwidth]{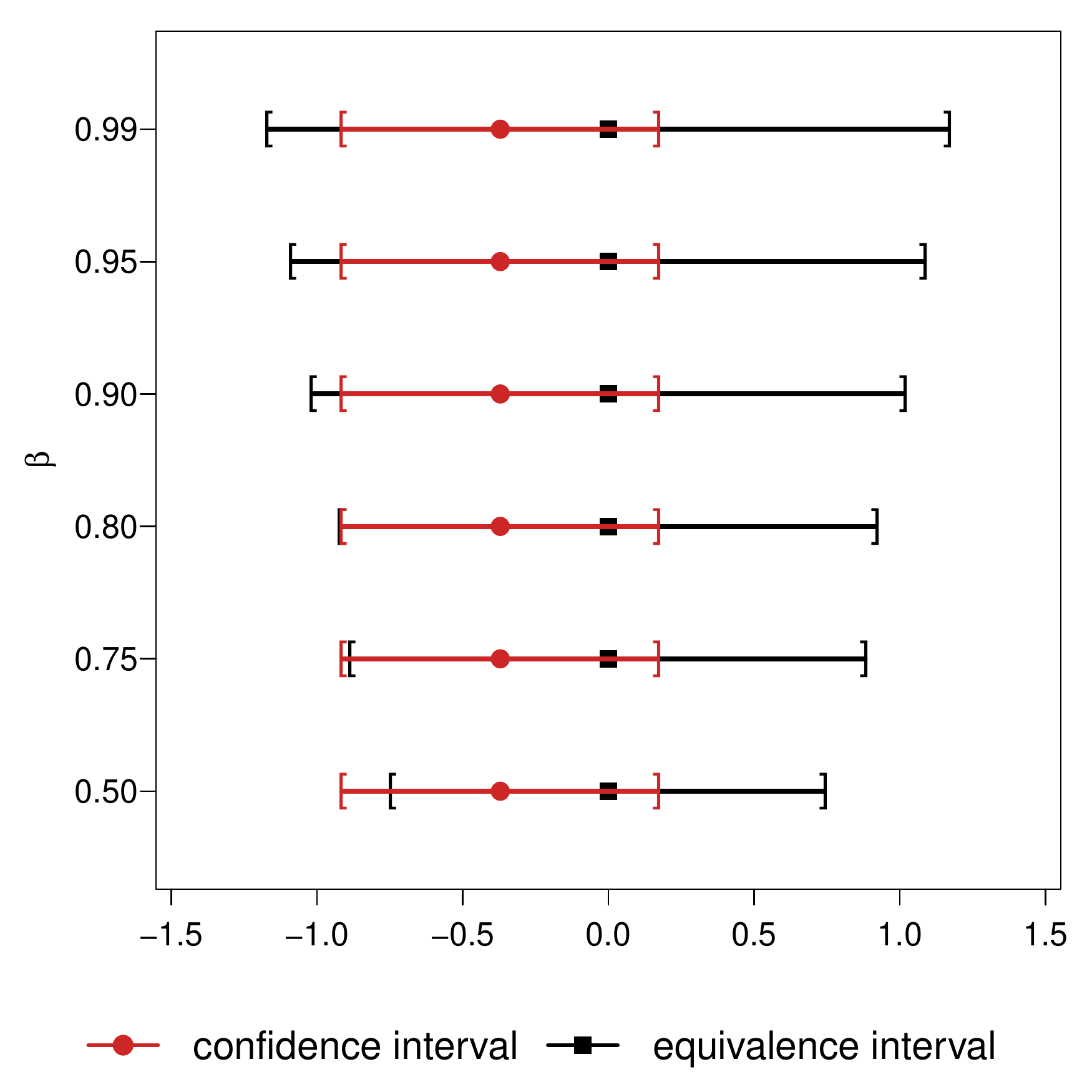}}

        \subfloat[marginal EEB of $\text{trt2}-\text{ctrl}$]{\includegraphics[width=0.45\textwidth]{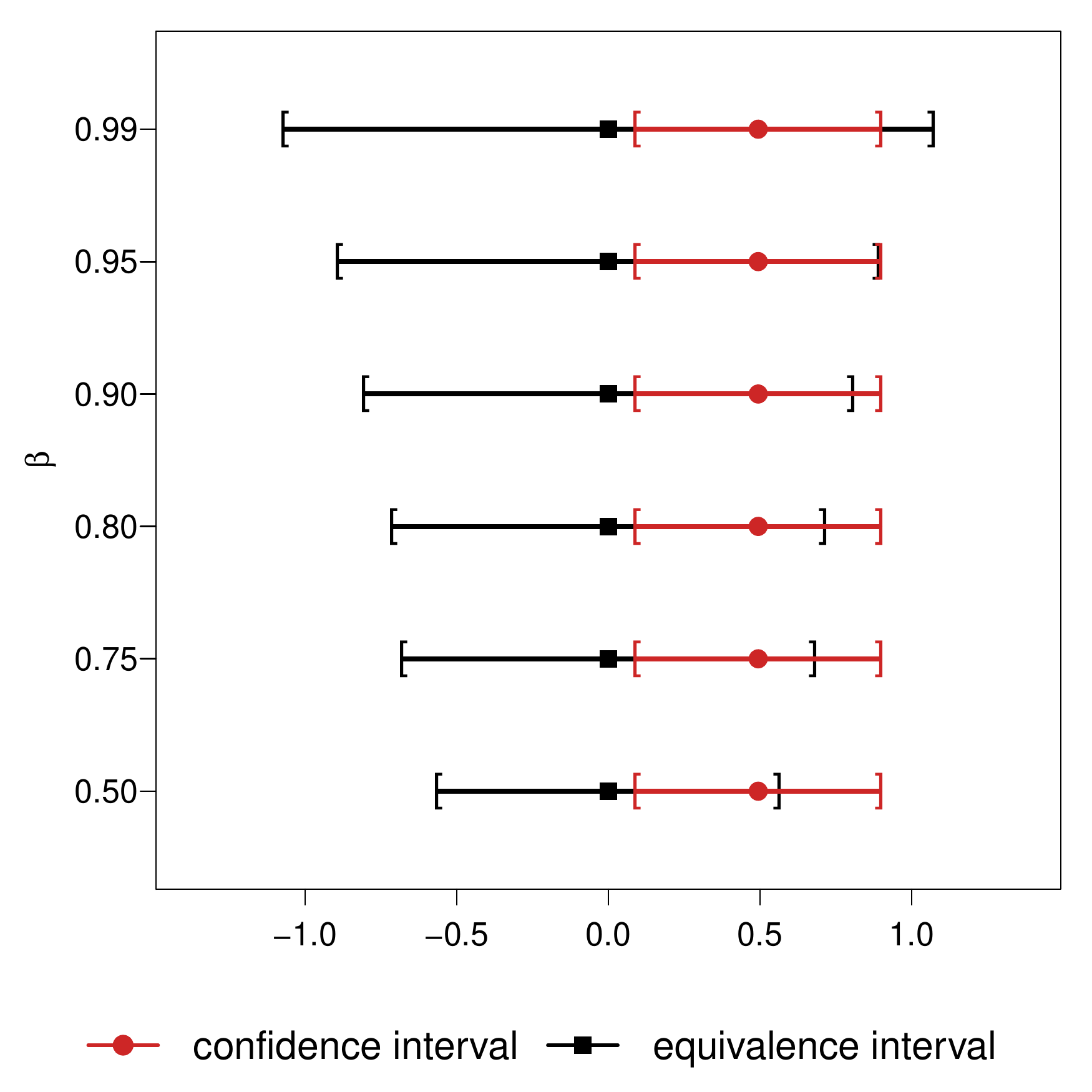}}
        \enskip{}
        \subfloat[conditional EEB of $\text{trt2}-\text{ctrl}$]{\includegraphics[width=0.45\textwidth]{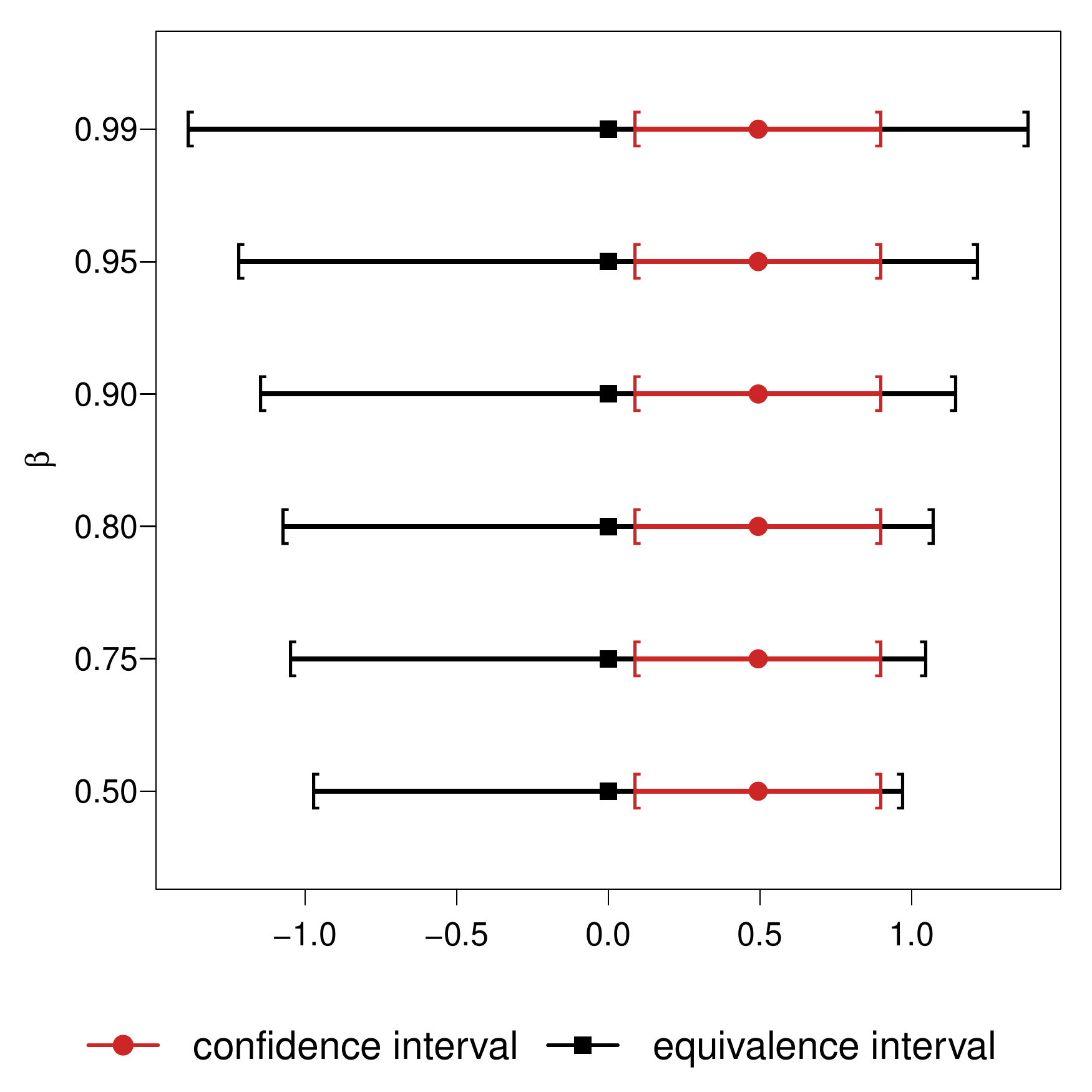}}
    \end{center}
    \caption{\label{fig:plant_EEB} The marginal and conditional equivalence interval at various $\beta$ levels for comparisons (a)\&(b) $\text{trt1}-\text{ctrl}$ and (c)\&(d) $\text{trt2}-\text{ctrl}$ in the Plant Growh study. The red intervals are the $90\%$ confidence interval, and the black ones are the equivalence intervals based on the marginal/conditional EEB at different $\beta$ level.}
\end{figure}

In this section, we use the Plant Growth Data Set to demonstrate the practical implementation of the proposed $B$-value and the EEB. The data were collected to compare yields obtained, which were measured by dried weight of plants, from a control and two distinct treatment groups~\citep{dobson1983introduction}. The data set is available in the open source software \textsf{R}~\citep{Rsoftware}. There are ten samples in each group. Figure~\ref{fig:plant_dist} shows the boxplot and distribution of the data, and Table~\ref{table:plant} presents the statistics in the two-sample $t$-test. From the table, it suggests that the weight in treatment group 2 is significantly higher than that in the control group; while there is no sufficient evidence showing that there is difference between treatment group 1 and control. Based on the testing results, we further conduct an equivalence test using the (conditional) EEB. Figure~\ref{fig:plant_EEB} displays both the marginal and conditional EEB for each comparison at difference $\beta$ level. For the comparison between treatment 1 and control, using the conditional equivalence interval, with $\beta$ level $\geq 0.8$, the equivalence test leads to a conclusion of equivalence, suggesting that there is no difference in dried plant weight between treatment 1 and control. Using the marginal EEB, the minimum $\beta$-level to conclude equivalence is about $0.75$. For the comparison between treatment 2 and control, though the two-sample $t$-test suggests a significant difference, with $\beta\geq0.5$, the equivalence interval derived from the conditional EEB fully covers the $90\%$ confidence interval, implying equivalence between the two groups. By running a second-stage equivalence test with the conditional EEB, we investigate the possibility that the original conclusion of a difference between treatment group 2 and the control group is possibly a false positive. With the marginal EEB, the equivalence test procedure is less efficient, since it needs $\beta$ to be as high as $0.99$.

\begin{table}
    \caption{\label{table:plant} Statistics of the two-sample $t$-test comparing the two treatment groups (trt1 and trt2) with the control group (ctrl) for the Plant Growth Data.}
    \begin{center}
        \begin{tabular}{l r r r r r}
            \hline
            \multicolumn{1}{c}{Comparison} & \multicolumn{1}{c}{Estimate (SE)} & \multicolumn{1}{c}{$t$-statistic} & \multicolumn{1}{c}{$p$-value} & \multicolumn{1}{c}{95\% CI} & \multicolumn{1}{c}{90\% CI} \\
            \hline
            $\text{trt1}-\text{ctrl}$ & -0.371 (0.311) & -1.191 & 0.249 & $[-1.025, 0.283]$ & $[-0.911, 0.169]$ \\
            $\text{trt2}-\text{ctrl}$ & 0.494 (0.231) & 2.134 & 0.047 & $[0.008, 0.980]$ & $[0.092, 0.895]$ \\
            \hline
        \end{tabular}
    \end{center}
\end{table}

\section{Discussion}
\label{sec:discussion}

The use of $P$-values in hypothesis testing has a long history, dating back to 1925 when R. A. Fisher introduced and promoted it for rejecting a null hypothesis when small~\citep{fisher1925statistical}: ``We shall not often be astray if we draw a conventional line at $0.05$.'' Since then, there are ongoing discussion about how to correctly use and interpret $P$-values. Though alternative statistics, for example confidence intervals, effect sizes or Bayes factors, are available, it has been shown that the interpretation of uncertainty is similar~\citep{wetzels2011statistical}. Common misunderstandings and misuse of $P$-values is likely partially responsible for general confusion and mistrust of empirical findings. In 2015, the editors of Basic and Applied Social Psychology (BASP) decided to ban $P$-values \cite[null hypothesis significance testing,][]{trafimow2015editorial}, a controversial move, even among opponents of null hypothesis significance testing and $P$-value usage. In 2016, the American Statistical Association announced a policy statement on $P$-values~\citep{wasserstein2016asa}. In the statement, the authors noted that ``the statistical community has been deeply concerned about issues of \emph{reproducibility} and \emph{replicability} of scientific conclusions''. In an echo to \citet{peng2015reproducibility} and \citet{leek2015statistics}, the statement also accentuated that ``misunderstanding and misuse of statistical inference is only one cause of the reproducibility crisis''. Hashing out opinions from more than two dozens of well-respect statisticians, the statement outlines six principles in order to regulate the proper use and interpretation of the $P$-values. One principle states that ``a $p$-value, or statistical significance, does not measure the size of an effect or the importance of a result'', as a $P$-value highly depends on the precision of the estimate (or the sample size). Recently, \citet{blume2018second} introduced the second-generation $P$-value and \citet{goodman2019proposed} proposed a hybrid effect size plus $P$-value criterion. Both criteria assemble the $P$-value (or confidence interval) with a practical/scientific significance in the testing procedure. However, same as in the equivalence test, the definition of practical/scientific significance can be subjective and arbitrary.

In this study, we introduced the B-value and the Empirical Equivalence Bound  and proposed a two-stage procedure when comparing two means from Gaussian distributed data. Our method is a data-driven procedure relaxing the knowledge of the equivalence level in an equivalence test. In the equivalence test, the conclusion highly depends on the nominal equivalence bound, which can be subjective. Our method eliminates this drawback by using the empirical equivalence bound derived from the data. On the other hand, performing a second-stage equivalence test also provides an opportunity to examine whether the significant result in the conventional two-sample $t$-test is a false positive discovery. This new two-stage testing procedure may then help improve the reproducibility in findings across studies.


\bibliographystyle{apalike}
\bibliography{main}

\clearpage

\appendix
\counterwithin{figure}{section}
\counterwithin{table}{section}
\counterwithin{equation}{section}


\section{Distribution of the $B$-value}
\label{appendix:sec:dist_B}

In this section, we derive the density function of $B$. Definition~\ref{def:dist_B} is a special scenario when the true parameter of interest $\delta$ is zero.

First, we derive the marginal cumulative distribution function of $B$. $B=\max\{|L|,|U|\}$, where 
\[
    L=\hat{\delta}-t_{\nu,1-\alpha}S, \quad U=\hat{\delta}+t_{\nu,1-\alpha}S,
\]
and the estimator $\hat{\delta}$ follows a Student's $t$-distribution with degrees of freedom $\nu$
\[
    \frac{\hat{\delta}-\delta}{S}\sim t(\nu),
\]
where $S$ is the standard error and $t_{\nu,1-\alpha}$ is the $100(1-\alpha)\%$ quantile of a $t$-distribution with degrees of freedom $\nu$. Thus, $[L,U]$ is the $100(1-2\alpha)\%$ confidence interval.

For $\forall~b\geq 0$, the cumulative distribution function of $B$ is
\begin{eqnarray*}
    F_{B}(b) &=& \mathbb{P}(B\leq b) \\
        &=& \mathbb{P}(|L|\leq b, B=|L|)+\mathbb{P}(|U|\leq b, B=|U|) \\
        &=& \mathbb{P}(|L|\leq b,L<-U<0<U)+\mathbb{P}(|L|\leq b,L<U<0) \\
        && +\mathbb{P}(|U|\leq b, L<0<-L<U)+\mathbb{P}(|U|\leq b, 0<L<U) \\
        &=& \mathbb{P}(-L\leq b, L<-U<0<U)+\mathbb{P}(-L\leq b,L<U<0) \\
        && +\mathbb{P}(U\leq b,L<0<-L<U)+\mathbb{P}(U\leq b,0<L<U).
\end{eqnarray*}
Denote $F_{t}(\cdot;\nu)$ as the cumulative distribution function of a Student's $t$-distribution with degrees of freedom $\nu$,
\begin{eqnarray*}
    \mathbb{P}(|L|\leq b,L<-U<0<U) &=& \mathbb{P}\left(\frac{\hat{\delta}-\delta}{S}\geq t_{\nu,1-\alpha}+\frac{-b-\delta}{S}  ,\frac{\hat{\delta}-\delta}{S}<-\frac{\delta}{S},\frac{\hat{\delta}-\delta}{S}>-t_{\nu,1-\alpha}-\frac{\delta}{S}\right) \\
    &=&\mathbb{P}\left(\max\left(t_{\nu,1-\alpha}+\frac{-b-\delta}{S},-t_{\nu,1-\alpha}-\frac{\delta}{S}\right)\leq\frac{\hat{\delta}-\delta}{S}<-\frac{\delta}{S}\right);
\end{eqnarray*}
\begin{eqnarray*}
    \mathbb{P}(|L|\leq b, L<U<0) &=& \mathbb{P}\left(\frac{\hat{\delta}-\delta}{S}\geq t_{\nu,1-\alpha}+\frac{-b-\delta}{S},\frac{\hat{\delta}-\delta}{S}<-t_{\nu,1-\alpha}-\frac{\delta}{S}\right) \\
    &=& \mathbb{P}\left(t_{\nu,1-\alpha}+\frac{-b-\delta}{S}\leq\frac{\hat{\delta}-\delta}{S}<-t_{\nu,1-\alpha}-\frac{\delta}{S}\right);
\end{eqnarray*}
\begin{eqnarray*}
    \mathbb{P}(|U|\leq b,L<0<-L<U) &=& \mathbb{P}\left(\frac{\hat{\delta}-\delta}{S}\leq -t_{\nu,1-\alpha}+\frac{b-\delta}{S}  ,\frac{\hat{\delta}-\delta}{S}>-\frac{\delta}{S},\frac{\hat{\delta}-\delta}{S}<t_{\nu,1-\alpha}-\frac{\delta}{S}\right) \\
    &=&\mathbb{P}\left(-\frac{\delta}{S}<\frac{\hat{\delta}-\delta}{S}\leq \min\left(-t_{\nu,1-\alpha}+\frac{b-\delta}{S},t_{\nu,1-\alpha}-\frac{\delta}{S}\right)\right);
\end{eqnarray*}
\begin{eqnarray*}
    \mathbb{P}(|U|\leq b, 0<L<U) &=& \mathbb{P}\left(\frac{\hat{\delta}-\delta}{S}\leq -t_{\nu,1-\alpha}+\frac{b-\delta}{S},\frac{\hat{\delta}-\delta}{S}>t_{\nu,1-\alpha}-\frac{\delta}{S}\right) \\
    &=& \mathbb{P}\left(t_{\nu,1-\alpha}-\frac{\delta}{S}<\frac{\hat{\delta}-\delta}{S}\leq -t_{\nu,1-\alpha}+\frac{b-\delta}{S}\right)
\end{eqnarray*}
Thus, if $St_{\nu,1-\alpha}\leq b<2St_{\nu,1-\alpha}$
\[
    F_{B}(b)=F_{t}\left(-t_{\nu,1-\alpha}+\frac{b-\delta}{S};\nu\right)-F_{t}\left(t_{\nu,1-\alpha}+\frac{-b-\delta}{S};\nu\right);
\]
and if if $b\geq 2St_{\nu,1-\alpha}$
\[
    F_{B}(b)=F_{t}\left(-t_{\nu,1-\alpha}+\frac{b-\delta}{S};\nu\right)-F_{t}\left(t_{\nu,1-\alpha}+\frac{-b-\delta}{S};\nu\right).
\]
Therefore, the cumulative function of $B$ is for $b\geq St_{\nu,1-\alpha}$
\begin{equation}\label{appendix:eq:cdf_B}
    F_{B}(b)=F_{t}\left(-t_{\nu,1-\alpha}+\frac{b-\delta}{S};\nu\right)-F_{t}\left(t_{\nu,1-\alpha}+\frac{-b-\delta}{S};\nu\right);
\end{equation}
and the probability density function is
\begin{equation}\label{appendix:eq:pdf_B}
    f_{B}(b) = \frac{\mathrm{d}F_{B}(b)}{\mathrm{d}b} =\frac{1}{S}\left\{f_{t}\left(-t_{\nu,1-\alpha}+\frac{b-\delta}{S};\nu\right)+f_{t}\left(t_{\nu,1-\alpha}+\frac{-b-\delta}{S};\nu\right)\right\},
\end{equation}
where $f_{t}(\cdot;\nu)$ is the probability density function of a Student's $t$-distribution with degrees of freedom $\nu$.

With a special case that $\delta=0$, function~\eqref{appendix:eq:cdf_B} degenerates to the distribution function in Definition~\ref{def:dist_B}.

Now, we consider the conditional distribution of $B$ given that $0\in[L_{0},U_{0}]$. 
\[
    F_{B}(b~|~0\in[L_{0},U_{0}]) = \mathbb{P}\left(B\leq b~|~0\in[L_{0},U_{0}]\right) = \frac{\mathbb{P}\left(B\leq b, 0\in[L_{0},U_{0}]\right)}{\mathbb{P}\left(0\in[L_{0},U_{0}]\right)}.
\]
\begin{eqnarray*}
    \mathbb{P}\left(0\in[L_{0},U_{0}]\right) &=& \mathbb{P}\left(\hat{\delta}-t_{\nu,1-\alpha/2}S<0<\hat{\delta}+t_{\nu,1-\alpha/2}S\right) \\
    &=& \mathbb{P}\left(\frac{\hat{\delta}-\delta}{S}<t_{\nu,1-\alpha/2}-\frac{\delta}{S},-t_{\nu,1-\alpha2}-\frac{\delta}{S}<\frac{\hat{\delta}-\delta}{S}\right) \\
    &=& F_{t}\left(t_{\nu,1-\alpha/2}-\frac{\delta}{S};\nu\right)-F_{t}\left(-t_{\nu,1-\alpha/2}-\frac{\delta}{S};\nu\right).
\end{eqnarray*}
\begin{eqnarray*}
    && \mathbb{P}\left(B\leq b, 0\in[L_{0},U_{0}]\right) \\
    &=& \mathbb{P}\left[B\leq b,-U_{0}<L_{0}<0<U_{0}\right]+\mathbb{P}\left[B\leq b, L_{0}<0<U_{0}<-L_{0}\right] \\
    &=& \mathbb{P}\left[U\leq b, -\hat{\delta}-t_{\nu,1-\alpha/2}S<\hat{\delta}-t_{\nu,1-\alpha/2}S<0<\hat{\delta}+t_{\nu,1-\alpha/2}S\right] \\
    && +\mathbb{P}\left[L\leq b, \hat{\delta}-t_{\nu,1-\alpha/2}S<0<\hat{\delta}+t_{\nu,1-\alpha/2}S<-\hat{\delta}+t_{\nu,1-\alpha/2}S \right] \\
    &=& \mathbb{P}\left[\hat{\delta}+t_{\nu,1-\alpha}S\leq b, \hat{\delta}>0,\hat{\delta}-t_{\nu,1-\alpha/2}S<0<\hat{\delta}+t_{\nu,1-\alpha/2}S\right] \\
    && +\mathbb{P}\left[-\hat{\delta}+t_{\nu,1-\alpha}S\leq b, \hat{\delta}<0,\hat{\delta}-t_{\nu,1-\alpha/2}S<0<\hat{\delta}+t_{\nu,1-\alpha/2}S\right] \\
    &=& \mathbb{P}\left[\frac{\hat{\delta}-\delta}{S}\leq\frac{b-\delta}{S}-t_{\nu,1-\alpha},\frac{\hat{\delta}-\delta}{S}>-\frac{\delta}{S},-t_{\nu,1-\alpha/2}-\frac{\delta}{S}<\frac{\hat{\delta}-\delta}{S}<t_{\nu,1-\alpha/2}-\frac{\delta}{S}\right] \\
    && +\mathbb{P}\left[\frac{\hat{\delta}-\delta}{S}\geq t_{\nu,1-\alpha}+\frac{-b-\delta}{S},\frac{\hat{\delta}-\delta}{S}<-\frac{\delta}{S},-t_{\nu,1-\alpha/2}-\frac{\delta}{S}<\frac{\hat{\delta}-\delta}{S}<t_{\nu,1-\alpha/2}-\frac{\delta}{S}\right] \\
    &=& \mathbb{P}\left[-\frac{\delta}{S}<\frac{\hat{\delta}-\delta}{S}\leq\min\left(\frac{b-\delta}{S}-t_{\nu,1-\alpha},t_{\nu,1-\alpha/2}-\frac{\delta}{S}\right)\right] \\
    &&+\mathbb{P}\left[\max\left(t_{\nu,1-\alpha}+\frac{-b-\delta}{S},-t_{\nu,1-\alpha/2}-\frac{\delta}{S}\right)<\frac{\hat{\delta}-\delta}{S}<-\frac{\delta}{S}\right] \\
    &=& F_{t}\left\{\min\left(\frac{b-\delta}{S}-t_{\nu,1-\alpha},t_{\nu,1-\alpha/2}-\frac{\delta}{S}\right);\nu\right\}-F_{t}\left\{\max\left(t_{\nu,1-\alpha}-\frac{b+\delta}{S},-t_{\nu,1-\alpha/2}-\frac{\delta}{S}\right);\nu\right\}.
\end{eqnarray*}
$\Rightarrow$
\begin{eqnarray*}
    && F_{B}(b~|~0\in[L_{0},U_{0}]) \\
    &=& \frac{F_{t}\left\{\min\left(\frac{b-\delta}{S}-t_{\nu,1-\alpha},t_{\nu,1-\alpha/2}-\frac{\delta}{S}\right);\nu\right\}-F_{t}\left\{\max\left(t_{\nu,1-\alpha}-\frac{b+\delta}{S},-t_{\nu,1-\alpha/2}-\frac{\delta}{S}\right);\nu\right\}}{F_{t}\left(t_{\nu,1-\alpha/2}-\frac{\delta}{S};\nu\right)-F_{t}\left(-t_{\nu,1-\alpha/2}-\frac{\delta}{S};\nu\right)}
\end{eqnarray*}
\begin{itemize}
  \item if $b\leq S(t_{\nu,1-\alpha}+t_{\nu,1-\alpha/2})$
 \[
    \begin{cases}
        \min\left(\frac{b-\delta}{S}-t_{\nu,1-\alpha},t_{\nu,1-\alpha/2}-\frac{\delta}{S}\right)=\frac{b-\delta}{S}-t_{\nu,1-\alpha} \\
        \max\left(t_{\nu,1-\alpha}-\frac{b+\delta}{S},-t_{\nu,1-\alpha/2}-\frac{\delta}{S}\right)=t_{\nu,1-\alpha}-\frac{b+\delta}{S}
    \end{cases}
 \]
 \[
    \frac{b-\delta}{S}-t_{\nu,1-\alpha}\geq t_{\nu,1-\alpha}-\frac{b+\delta}{S} \quad \Rightarrow \quad b\geq St_{\nu,1-\alpha}
 \]
   \item if $b> S(t_{\nu,1-\alpha}+t_{\nu,1-\alpha/2})$
  \[
    \begin{cases}
        \min\left(\frac{b-\delta}{S}-t_{\nu,1-\alpha},t_{\nu,1-\alpha/2}-\frac{\delta}{S}\right)=t_{\nu,1-\alpha/2}-\frac{\delta}{S} \\
        \max\left(t_{\nu,1-\alpha}-\frac{b+\delta}{S},-t_{\nu,1-\alpha/2}-\frac{\delta}{S}\right)=-t_{\nu,1-\alpha/2}-\frac{\delta}{S}
    \end{cases}
 \]
 \[
    t_{\nu,1-\alpha/2}-\frac{\delta}{S} > -t_{\nu,1-\alpha/2}-\frac{\delta}{S} \quad (\Rightarrow \quad F_{B}(b~|~0\in[L_{0},U_{0}])=1)
 \]
\end{itemize}
Therefore,
\begin{itemize}
    \item $b\in(-\infty,St_{\nu,1-\alpha})$
        \[
            F_{B}(b~|~0\in[L_{0},U_{0}])=0
        \]
  \item $b\in [St_{\nu,1-\alpha},S(t_{\nu,1-\alpha}+t_{\nu,1-\alpha/2})]$
    \[
        F_{B}(b~|~0\in[L_{0},U_{0}])=\frac{F_{t}((b-\delta)/S-t_{\nu,1-\alpha};\nu)-F_{t}(t_{\nu,1-\alpha}-(b+\delta)/S;\nu)}{F_{t}(t_{\nu,1-\alpha/2}-\delta/S;\nu)-F_{t}(-t_{\nu,1-\alpha/2}-\delta/S;\nu)}
    \]
  \item $b\in(S(t_{\nu,1-\alpha}+t_{\nu,1-\alpha2}),\infty)$ 
    \[
        F_{B}(b~|~0\in[L_{0},U_{0}])=1
    \]
\end{itemize}

Analogously, for the conditional distribution of $B$ given that $0\notin[L_{0},U_{0}]$,
\[
    F_{B}(b~|~0\notin[L_{0},U_{0}]) = \mathbb{P}\left(B\leq b~|~0\notin[L_{0},U_{0}]\right) = \frac{\mathbb{P}\left(B\leq b, 0\notin[L_{0},U_{0}]\right)}{\mathbb{P}\left(0\notin[L_{0},U_{0}]\right)}.
\]
\begin{eqnarray*}
    \mathbb{P}\left(0\notin[L_{0},U_{0}]\right) &=& \mathbb{P}\left(0<L_{0}<U_{0}\right)+\mathbb{P}\left(L_{0}<U_{0}<0\right) \\
    &=& \mathbb{P}\left(0<\hat{\delta}-t_{\nu,1-\alpha/2}S\right)+\mathbb{P}\left(\hat{\delta}+t_{\nu,1-\alpha/2}S<0\right) \\
    &=& \mathbb{P}\left(t_{\nu,1-\alpha/2}-\frac{\delta}{S}<\frac{\hat{\delta}-\delta}{S}\right)+\mathbb{P}\left(\frac{\hat{\delta}-\delta}{S}<-t_{\nu,1-\alpha/2}-\frac{\delta}{S}\right) \\
    &=& \left[1-F_{t}\left(t_{\nu,1-\alpha/2}-\frac{\delta}{S};\nu\right)\right]+F_{t}\left(-t_{\nu,1-\alpha/2}-\frac{\delta}{S};\nu\right) \\
    &=& F_{t}\left(\frac{\delta}{S}-t_{\nu,1-\alpha/2};\nu\right)+F_{t}\left(-t_{\nu,1-\alpha/2}-\frac{\delta}{S};\nu\right)
\end{eqnarray*}
\begin{eqnarray*}
    && \mathbb{P}\left(B\leq b, 0\notin[L_{0},U_{0}]\right) \\
    &=& \mathbb{P}\left(B\leq b, 0<L_{0}<U_{0}\right)+\mathbb{P}\left(B\leq b, L_{0}<U_{0}<0\right) \\
    &=& \mathbb{P}\left(U\leq b, 0<L_{0}<U_{0}\right)+\mathbb{P}\left(-L\leq b, L_{0}<U_{0}<0\right) \\
    &=& \mathbb{P}\left(\hat{\delta}+t_{\nu,1-\alpha}S\leq b, \hat{\delta}-t_{\nu,1-\alpha/2}S>0\right)+\mathbb{P}\left(-\hat{\delta}+t_{\nu,1-\alpha}S\leq b, \hat{\delta}+t_{\nu,1-\alpha/2}S<0\right) \\
    &=& \mathbb{P}\left(t_{\nu,1-\alpha/2}-\frac{\delta}{S}<\frac{\hat{\delta}-\delta}{S}\leq\frac{b-\delta}{S}-t_{\nu,1-\alpha}\right)+\mathbb{P}\left(t_{\nu,1-\alpha}+\frac{-b-\delta}{S}\leq\frac{\hat{\delta}-\delta}{S}<-t_{\nu,1-\alpha/2}-\frac{\delta}{S}\right) \\
    &=& \left[F_{t}\left(\frac{b-\delta}{S}-t_{\nu,1-\alpha};\nu\right)-F_{t}\left(t_{\nu,1-\alpha/2}-\frac{\delta}{S};\nu\right)\right]+\left[F_{t}\left(-t_{\nu,1-\alpha/2}-\frac{\delta}{S};\nu\right)-F_{t}\left(t_{\nu,1-\alpha}+\frac{-b-\delta}{S};\nu\right)\right]
\end{eqnarray*}
$\Rightarrow$
\begin{itemize}
    \item $b\in(-\infty,S(t_{\nu,1-\alpha}+t_{\nu,1-\alpha/2}))$
        \[
            F_{B}(b~|~0\notin[L_{0},U_{0}])=0
        \]
    \item $b\in [S(t_{\nu,1-\alpha}+t_{\nu,1-\alpha/2}),\infty)$
        \begin{eqnarray*}
            && F_{B}(b~|~0\notin[L_{0},U_{0}]) \\
            &=& \frac{\left[F_{t}\left(\frac{b-\delta}{S}-t_{\nu,1-\alpha};\nu\right)-F_{t}\left(t_{\nu,1-\alpha/2}-\frac{\delta}{S};\nu\right)\right]+\left[F_{t}\left(-t_{\nu,1-\alpha/2}-\frac{\delta}{S};\nu\right)-F_{t}\left(t_{\nu,1-\alpha}+\frac{-b-\delta}{S};\nu\right)\right]}{F_{t}\left(\frac{\delta}{S}-t_{\nu,1-\alpha/2};\nu\right)+F_{t}\left(-t_{\nu,1-\alpha/2}-\frac{\delta}{S};\nu\right)}
        \end{eqnarray*}
\end{itemize}

When $\delta=0$, the (conditional) distribution functions have the same form as in Definition~\ref{def:dist_B}.

\section{Calculation of the Empirical Equivalence Bound}
\label{appendix:sec:EEB}

Following Definitions~\ref{def:dist_B} and \ref{def:EEB}, we can calculate the Empirical Equivalence Bound (EEB) from the data. When the true difference $\delta=0$ and under conditions $0\in[L_{0},U_{0}]$ and $0\notin[L_{0},U_{0}]$, EEB has explicit forms as in Section~\ref{sub:method_formulation}. For other scenarios, a numerical solution of EEB can be obtained by using the bisection method.




\end{document}